\documentclass[journal]{IEEEtran}

\usepackage{float}
\usepackage{graphicx,subfigure}
\usepackage{bm}
\usepackage{amsmath}
\usepackage{amssymb}
\usepackage{latexsym}
\usepackage{epsfig}
\usepackage{amsbsy}
\usepackage{array}
\usepackage{amssymb}
\usepackage{booktabs}
\usepackage{bbding}
\usepackage{multirow}
\usepackage{slashbox}
\usepackage{mathrsfs}

\usepackage[ruled,norelsize]{algorithm2e}

\usepackage{graphicx}
\graphicspath{{fig_pdf/}}
\newlength\figwidth
\usepackage[mathlines]{lineno}

\ifCLASSINFOpdf
\else
\fi

\usepackage{hyperref}
\hypersetup{
linktocpage=true,
pdfborderstyle={/S/S/W 1},
hyperindex=true,
bookmarks=true,
bookmarksopen=true,
bookmarksnumbered=true,
}

\newcommand{\vct}[1]{\bm{#1}}
\newcommand{\mtx}[1]{\bm{#1}}

\newcommand{\Fee}{\mtx{\Phi}}

\newcommand{\norm}[1]{\left\Vert {#1} \right\Vert}
\newcommand{\pnorm}[2]{\norm{#2}_{#1}}

\makeatletter
\newcommand{\removelatexerror}{\let\@latex@error\@gobble}
\makeatother

\begin{document}

\title{Greedy Sparse Signal Reconstruction Using Matching Pursuit Based on Hope-tree}

\author{Zhetao Li, Hongqing Zeng, Chengqing Li, and Jun Fang
\thanks{This work was supported by the National Natural Science Foundation of China (no. 61532020, 61379115).}
\thanks{Z. Li, H. Zeng, and C. Li are with the College of Information Engineering, Xiangtan University, Xiangtan 411105, Hunan, China (e-mail: DrChengqingLi@gmail.com).}
\thanks{J. Fang is with the National Key Laboratory of Science and Technology on Communications, University of Electronic Science and Technology of China,
Chengdu 611731, China (e-mail: JunFang@uestc.edu.cn).}
}

\markboth{IEEE Signal Processing Letters}%
{Li \MakeLowercase{\textit{et al.}}: }


\IEEEpubid{\begin{minipage}{\textwidth}\ \\[12pt] \centering
  1549-8328 \copyright 2016 IEEE. Personal use is permitted, but republication/redistribution requires IEEE permission.\\
  See http://www.ieee.org/publications standards/publications/rights/index.html for more information.
\end{minipage}}

\maketitle

\begin{abstract}
The reconstruction of sparse signals requires the solution of an $\ell_0$-norm minimization problem in Compressed Sensing. Previous research has focused on the investigation of a single candidate to identify the support (index of nonzero elements) of a sparse signal. To ensure that the optimal candidate can be obtained in each iteration, we propose here an iterative greedy reconstruction algorithm (GSRA). First, the intersection of the support sets estimated by the Orthogonal Matching Pursuit (OMP) and Subspace Pursuit (SP) is set as the initial support set. Then, a hope-tree is built to expand the set. Finally, a developed decreasing subspace pursuit method is used to rectify the candidate set. Detailed simulation results demonstrate that GSRA is more accurate
than other typical methods in recovering Gaussian signals, 0--1 sparse signals, and synthetic signals.
\end{abstract}

\IEEEpeerreviewmaketitle

\section{Introduction}

\IEEEPARstart{I}{n} the past decade, Compressed Sensing (CS) has received considerable attention as a means to reconstruct sparse signals in underdetermined systems. The basic premise of CS is that a $K$-sparse signal $\vct{x} \in  \mathbb{R}^{n}$ can be recovered perfectly with far fewer compressed measurements $\vct{y}=\Fee \vct{x} \in \mathbb{R}^m$ than the traditional approaches \cite{Tao:uncertain:TIT06,Terence:decode:TIT05,Shen:analysis:SPL13,Determe:condition:SPL16}. The problem of reconstructing the original signal can be formulated as an $\ell_0$-minimization problem
\begin{equation}
\mathop {\min }\limits_x \pnorm{0}{x} \quad \mbox{subject to} \quad \vct{y}=\Fee \vct{x},
\label{eq:1}
\end{equation}
where $\pnorm{p}{\bullet}$ denotes the $p$th norm, $\Fee \in \mathbb{R}^{m \times n}$ is the sensing matrix, and $m<n$. Since the $\ell_0$-minimization problem is NP-hard \cite{Tao:uncertain:TIT06}, it is often transformed into an $\ell_1$-minimization problem to make it tractable. Concretely, some $\ell_1$-relaxation methods, such as Basis Pursuit (BP) and BP denoising (BPDN), can be used to solve it \cite{Chen:atomic:SR01}.

To greatly reduce the computational complexity of the $\ell_1$-relaxation methods, algorithms based on greedy pursuit have been adopted, which use the least squares method to estimate the sparse signal $\vct{x}$ by finding the positions of its nonzero elements. The sparse signal can be represented as $\vct{x}=\sum_{j \in S}\vct{c}_j\vct{e}_j$, where the $\vct{e}_j$ are the standard unit vectors and $S$ is the support set of $\vct{x}$. In Orthogonal Matching Pursuit (OMP), the index of the column possessing the strongest correlation with the modified measurements is chosen as the new element of the support
in each iteration \cite{Wang:recovery:TSP12}. Note that if any index in the support set is not right, the result of OMP would be wrong. To mitigate this weakness of OMP, improved selection methods for indices have been developed, such as Stagewise Orthogonal Matching Pursuit (StOMP) \cite{Donoho:sparse:TIT12}, which adopts a threshold mechanism. Subspace pursuit (SP) and Compressive Sampling Matching Pursuit (CoSaMP) select indices exceeding the sparsity level by a pruning \cite{Dai:subspace:TIT09,Tropp:iterative:ITA}. Multipath Matching Pursuit (MMP) builds a tree to find the optimal support set \cite{Kwon:multipath:TIT14}. Generalized Orthogonal Matching Pursuit (gOMP) selects multiple indices \cite{Wang:generalized:TSP12}. However, these algorithms cannot obtain an acceptable trade-off behavior between computational complexity and reconstruction performance.

\IEEEpubidadjcol 

In this letter, we propose an iterative greedy signal reconstruction algorithm (GSRA) using matching pursuit based on hope-tree.
First, the intersection of the support sets is estimated by OMP and SP and taken as the initial support set.
Then, a hope-tree is built to obtain a candidate support $S$ by setting the search depth, which is rectified by a decreasing subspace pursuit method.
Finally, we calculate that the complexity of GSRA is $O(N_{\rm max}\cdot \mathit{mn}\cdot \mathit{iter})$, where $\mathit{iter}$ is the number of iterations and $N_{\rm max}$ is the number of candidates in each iteration.

The rest of this paper is organized as follows. In Sec.~\ref{sec:matching}, we introduce the matching pursuit used in the GSRA. In Sec.~\ref{sec:theory}, we provide the efficiency analysis of the GSRA. Section~\ref{sec:result} presents detailed empirical experiments on its reconstruction performance. The last section concludes the paper.

\section{MATCHING PURSUIT BASED ON HOPE-TREE}
\label{sec:matching}

The proposed GSRA is composed of three stages: pre-selection, hope-tree search, and rectification. These are described in the following three subsections.

\subsection{Pre-selection}

The purpose of pre-selection is to obtain the indices belonging to the target support set $T$ with a high probability. The initial support set $\Theta$ can be taken to be the intersection of the support sets estimated by different types of reconstruction algorithms, such as OMP, SP and gOMP. If three or more such
are used, the reconstruction performance cannot be improved further, but the whole complexity is greatly increased. As the index selection strategy of OMP is different from the others, we only adopt OMP and SP, so as to satisfy the computational efficiency requirement.

\subsection{Hope-Tree search}

\begin{figure}[!htb]
\centering
\begin{minipage}{\figwidth}
\centering
\includegraphics[width=\figwidth]{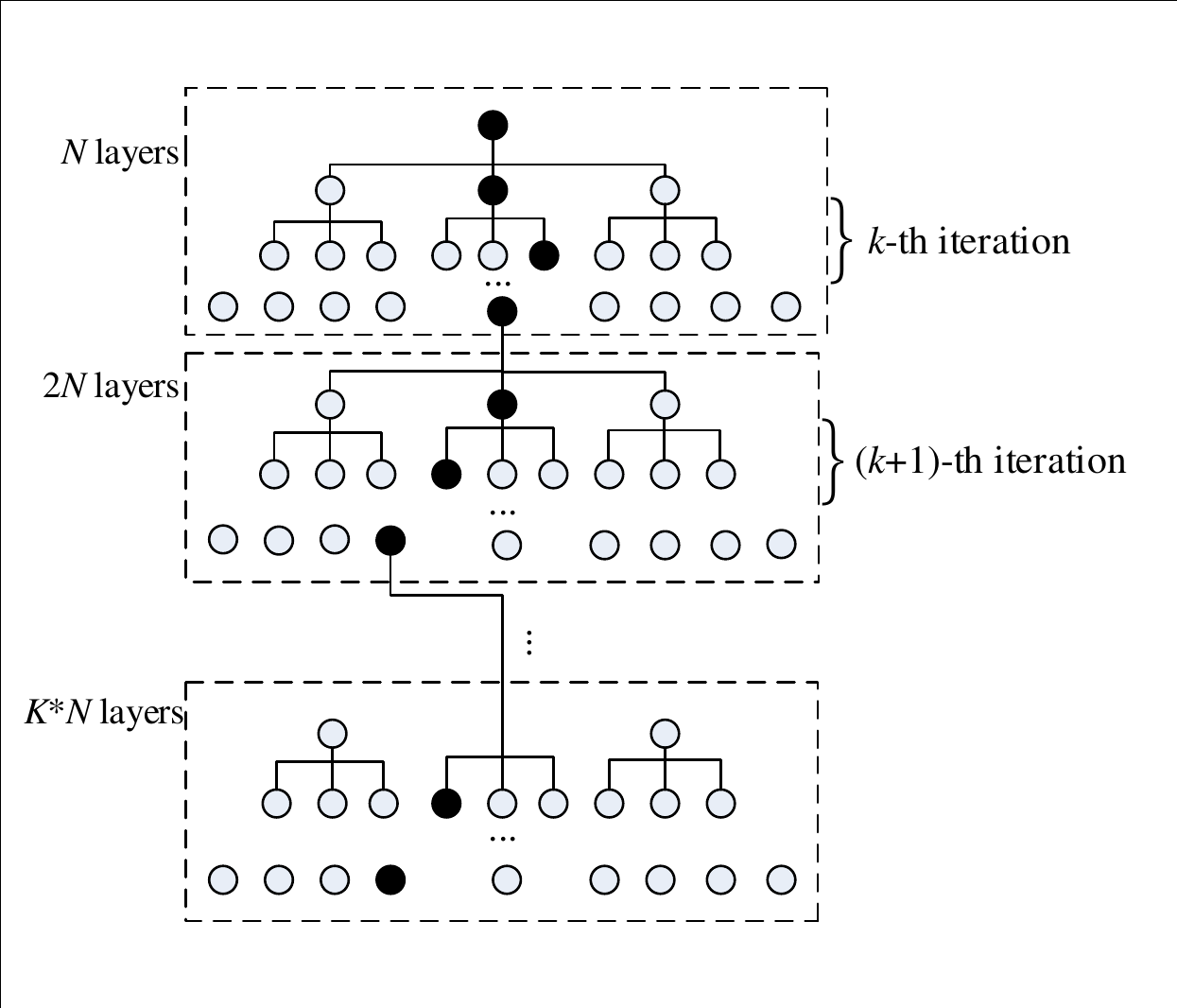}
\end{minipage}
\caption{Illustration of the hope-tree search, where the black dots denote the optimal candidate.}
\label{fig:Descript}
\end{figure}

Once the pre-selection is finished, a hope-tree is built to search for the candidate support. As shown in Fig.~\ref{fig:Descript}, the search depth is set as $N$ in the $k$th iteration, and each path generates $L\ (\le K)$ child paths with indices $\widetilde \pi, \widetilde \pi, \cdots, \widetilde \pi _L$ of those columns possessing the strongest correlations with the residual, according to the condition
\begin{equation}
\{\widetilde\pi_1, \widetilde\pi_2, \cdots, \widetilde\pi_L \} = \mathop{\arg\max }\limits_{|\pi|=L}\pnorm{2}{(\Fee^T\vct{r_i}^{k-1})_{\pi}}^2.
\label{eq:2}
\end{equation}
When the search depth approaches $N$, a sub-tree of the hope-tree is built. The modulo strategy in \cite{Kwon:multipath:TIT14} is used to compute the layer order, and execute a priority search for promising candidates. Then, the indices of the optimal paths with the minimum residuals are selected as the best elements of $T$ and added to the set $\Theta$. Next, the last optimal index of the $k$th iteration is chosen as the root index.
The above procedures are repeated in the $(k+1)$th iteration. The tree search procedure is stopped when the length of the support $\Theta$ is not smaller than $K$. In each sub-process, the number of candidates increases by the factor $L$, resulting in $L^N$ candidates after $N$ iterations. These operations are repeated until either the candidate support satisfies $S=\Theta$ or the $\ell_2$-norm of the residual falls below a preset threshold, i.e., $\pnorm{2}{\vct{r}^k}\le \varsigma$.

\subsection{Rectification}

Because searching the hope-tree may ignore any correct indices falling outside of the hope-tree, a developed decreasing subspace pursuit method is used to obtain the indices from the latter and improve the accuracy of the candidate support set.

In the first iteration, the size of the search space is $T=K$, and the decreasing subspace pursuit method expands the selected support with the indices of the $T$ largest magnitude entries of $\Fee^Tr$. Once the size of the extended support $S^k$ reaches $T+K$, the orthogonal projection coefficients of $\vct{y}$ onto $\Fee_{S^k}$ are calculated, and the estimated support $S^k$ is obtained by pruning it to contain only the $K$ indices corresponding to the largest elements of $\Fee_{S^k}^\dag \vct{y}$, where $k$ is the iteration index. In the next iteration, the length of the search space decreases according to $T_{k+1}=\alpha T_k$, while the execution remains the same as in the first iteration, where $0<\alpha<1$. When $T$ becomes smaller than a certain
threshold, the iteration is terminated. The final support set $S$ and the estimated version of the signal $\vct{x}$ are determined.

The pseudocode of GSRA is summarized as Algorithm~\ref{algo:mp}, where the process of building the hope-tree and
the subprocess of rectifying the candidate support set are summarized as Algorithm~\ref{algo:mmp} and Algorithm~\ref{algo:rec}, respectively.
The main merit of the reconstruction algorithm of GSRA is that it creates a process of searching the hope-tree and rectifying the candidate support set, which can balance the contradiction between computational complexity and reconstruction performance.

\removelatexerror
\begin{algorithm}[H]
\SetAlgoLined
\KwData{measurement vector $y$, Sensing Matrix $\Fee$, sparsity level $K$, number of path $L$, search depth $N $ \\
  $J_{\rm omp} = \mathit{OMP}(\Fee ,\vct{y},K)$, \quad // OMP support set \\
  $J_{\rm sp} = \mathit{SP}(\Fee ,\vct{y},K)$, \quad \quad// SP support set \\
 initial candidate support set $\lambda=J_{\rm omp}\cap J_{\rm sp}$.}
\KwResult{$\vct{\widehat x}$, $S$}
 {\bf Initialization:}\\
  Residual vector ${r^0} = \vct{y} - \Fee {\vct{x}_\lambda }$, candidate support ${\Lambda ^0} = \lambda$,
  backtracking subspace parameter $\alpha$ ($0<\alpha<1$), iteration index $k = 0 $ \\

  \While{${\| \Lambda \|_0} < K$}{
   $k=k+1$ \\
   $[\rm{supp}, \sim]$ = \rm{CreateHopeTree}$(\vct{y}, \Fee, \vct{r}^k, K, N, L)$\\
   {\bf Update candidate support:} ${\Lambda ^k} = \Lambda ^{k-1} \cup {\mathop{\rm supp}\nolimits}$ \\
   {\bf Estimate $\vct{\widehat x}_{\Lambda ^k}$:} $\vct{\widehat x}_{\Lambda ^k}= \Fee _{\Lambda ^k}^\dag\vct{y}$\\
   {\bf Update residual:} $\vct{r}^k=\vct{y}-\Fee_\Lambda \vct{\widehat x_{\Lambda ^k}}$
 }
[$\vct{\widehat x}$, $S$]=\rm{RectifySupport}$(\Lambda, \vct{y}, \Fee, \vct{r}_{\rm new}, \alpha)$
\caption{Matching Pursuit Based on Hope-Tree.}
\label{algo:mp}
\end{algorithm}

\begin{algorithm}
  \SetAlgoLined
  \KwData{Initialization: maximum number of search candidate $N_{\rm max}$, stop threshold $\varsigma$, candidate order $\ell = 0$, minimal magnitude of residual $\rho = \infty$}
  \KwResult{$supp$}
  \While{$\ell<N_{\rm max}$ \mbox{and} $\varsigma<\rho$}
  {
    $\ell=\ell+1$, $\vct{r^0}=\vct{r}$, $\mathit{temp}=\ell-1$ \\

    \For{$k=1$ to $N$}
    {
        $c_{k} = temp \bmod L+1$ // compute layer order\\
        $temp =\lfloor temp/L \rfloor$\\
        $c_k=c_k\cup c_{k-1}$
    }
    \For{$k=1$ to $N$}
    {
        $\widetilde \pi = \arg \mathop {\max }\limits_{| \pi | = L} \| ({\Fee ^T}\vct{r}^{k - 1})_\pi \|_2^2$    // choose $L$ best indices \\
        ${s_{\ell}^k}=s_{\ell}^{k-1} \cup \widehat {\pi}_{c_k}$\\
        $\vct{\widehat x}^k = \Fee_{s_\ell^k}^\dag\vct{y}$ \\
        $\vct{r}^k=\vct{y} - \Fee_{s_\ell^k}\vct{\widehat x}^k$
    }
    \lIf{$\|\vct{r}^N\| \le \rho$}
    {
        $supp = s$,\\ $\rho=\|\vct{r}^N\|$
    }
}
\caption{[$supp$, $\sim$]=\rm{CreateHopeTree}$(\vct{y}, \Fee, \vct{r}, K, N, L)$}
\label{algo:mmp}
\end{algorithm}

\begin{algorithm}
  \SetAlgoLined
  \KwData{the support set $\Lambda$, backtrack subspace parameter $T\ (T \le K)$ and residual $\vct{r}_{\rm new}$}
  \KwResult{$\vct x$ (estimated signal), $S$ (estimated support)}
  \While{$T$ $!=0$}{$S = $ {$T$ indices of highest amplitude components of $\Fee ^T r_{\rm new}$},
   $\Lambda=\Lambda\cup S$; $\vct{x} = A_\Lambda^\dag \vct{y}$ \\
   $S=K$ indices corresponding to the largest magnitude components of $\vct{x}$,
   $\vct{\widehat x}=\Fee_S^\dag \vct{y} $; $r_{\rm new}=\vct{y}-\Fee \vct{\widehat x} $; $T=\alpha T$}
\caption{[$\vct{\widehat x}$, $S$]=\rm{RectifySupport}$(\Lambda, \vct{y}, \Fee, \vct{r}_{\rm new}, \alpha)$}
\label{algo:rec}
\end{algorithm}

\section{EFFICIENCY ANALYSIS}
\label{sec:theory}

\subsection{Parameter setting}

A sensing matrix $\Fee$ is said to satisfy the restricted isometry property (RIP) of order $K$ if there exists a constant $\delta\in [0,1]$ such that
\begin{equation}
(1 - \delta)\pnorm{2}{ \vct{x}}^2 \leq \pnorm{2}{\vct{x}}^2 \leq (1 + \delta)\pnorm{2}{ \vct{x}}^2
\label{eq:3}
\end{equation}
for any $K$-sparse vector $\vct{x}$. In particular, the minimum of all constants $\delta$ ensuring that the correct indices can be obtained
is referred to as the isometry constant $\delta_K$. In the clean case, it is assumed that $\Fee$ satisfies the RIP of order $\mathit{NK}$, i.e., $\delta _{\rm NK} \in (0,1)$. So, one has $0<1-\delta _{\rm NK} \le \lambda_{\min}(\Fee_D^T\Fee _D)$ for any index set $N$ satisfying $|D|\le \mathit{NK}$, which indicates that all eigenvalues of $\Fee_D$ are positive. Thus, $\Fee_D$ has full column rank (i.e., $K \le m/N $) and $N\le \sqrt m$. Therefore, one has $N<\min(K, \sqrt m)$. However, in the noisy case, a noisy disturbance makes the GSRA improve its performance at the cost of increasing the value of the search depth. To make $\Fee$ satisfy RIP and $S$ include the correct indices in each optimal path as much as possible, the value of $N$ can be set appropriately with the upper bound $K$.

\subsection{Computational Complexity Analysis}

The complexities of the three stages of GSRA can be estimated separately. The complexity for obtaining the initial support set is
$$O_{\rm omp}( \mathit{mn} \cdot K )+O_{\rm sp}(\mathit{mn}\cdot K) \approx O(\mathit{mn}\cdot K),$$
where $O_{\rm omp}(\mathit{mn}\cdot K)$ and $O_{\rm sp}(\mathit{mn}\cdot K)$ are derived in \cite{Dai:subspace:TIT09}. The number of candidates in the $k$th iteration
is $L^N$. To decrease the computational complexity, the maximum number of candidates $N_{\rm max}$ is less than $L^N$. Obtaining one index requires $O(mn)$ according to the expression $\Fee^T\vct{r}^k$. Choosing $N$ indices from $N_{\rm max}$ candidates in each iteration means that the complexity of searching the hope-tree is $O(N_{\rm max}\cdot \mathit{mn})$. As the number of iterations is set to be $\mathit{iter}<K$, the complexity of searching the hope-tree is bounded by $O(N_{\rm max}\cdot\mathit{mn}\cdot \mathit{iter})$. In addition, the number of selecting elements is reduced in each iteration; the complexity of the decreasing subspace method is bounded by $O(mn)$. Therefore, the total complexity of GSRA is
$O(N_{\rm max}\cdot n\cdot \mathit{iter}) + O(\mathit{mn}\cdot K) + O(mn) \approx O(N_{\rm max}\cdot mn\cdot \mathit{iter})$. When the value of $K$ is low, the complexity of GSRA is close to that of OMP. In addition, the complexity of GSRA monotonously increases with respect to $K$ and far less than that of the breadth-first search method.

\section{SIMULATION RESULTS}
\label{sec:result}

To check the real performance of GSRA for solving the sparse signal recovery problem, it was compared with OMP, gOMP, MMP-DFS, MMP-BFS using the same setup adopted in \cite{Dai:subspace:TIT09}.

\subsection{Reconstruction of Sparse Signals Whose Nonzero Values Follow $\mathcal{N}(0,1)$}
\label{ssec:n01}

In this experiment, some important parameters were set as follows:
\begin{enumerate}
\item Generate a random matrix $\Fee$ of size $m \times n$ with i.i.d. $\mathcal N(0,1)$ entries. To facilitate comparison, we set $m=128$ and $n=256$.

\item To reduce the time complexity, set the maximum number of candidates of sub-tree $N_{\rm max}=30$, the number of paths $L=2$, and the search depth $N=7$.

\item Choose a $K$-subset of $\{1, 2, \cdots, n\}$, $K=1, 2, \cdots, 60$.

\item Set the value of $\vct{x}$ in the $K$ chosen indices as random nonzero values obeying $\mathcal{N}(0, 1)$ and that in the remaining indices as zero.

\item Conduct $1,000$ simulations to calculate the frequency of the exact reconstruction and the mean time needed for the different algorithms.
\end{enumerate}

\begin{figure}[!htb]
\centering
\begin{minipage}{\figwidth}
\centering
\includegraphics[width=\textwidth]{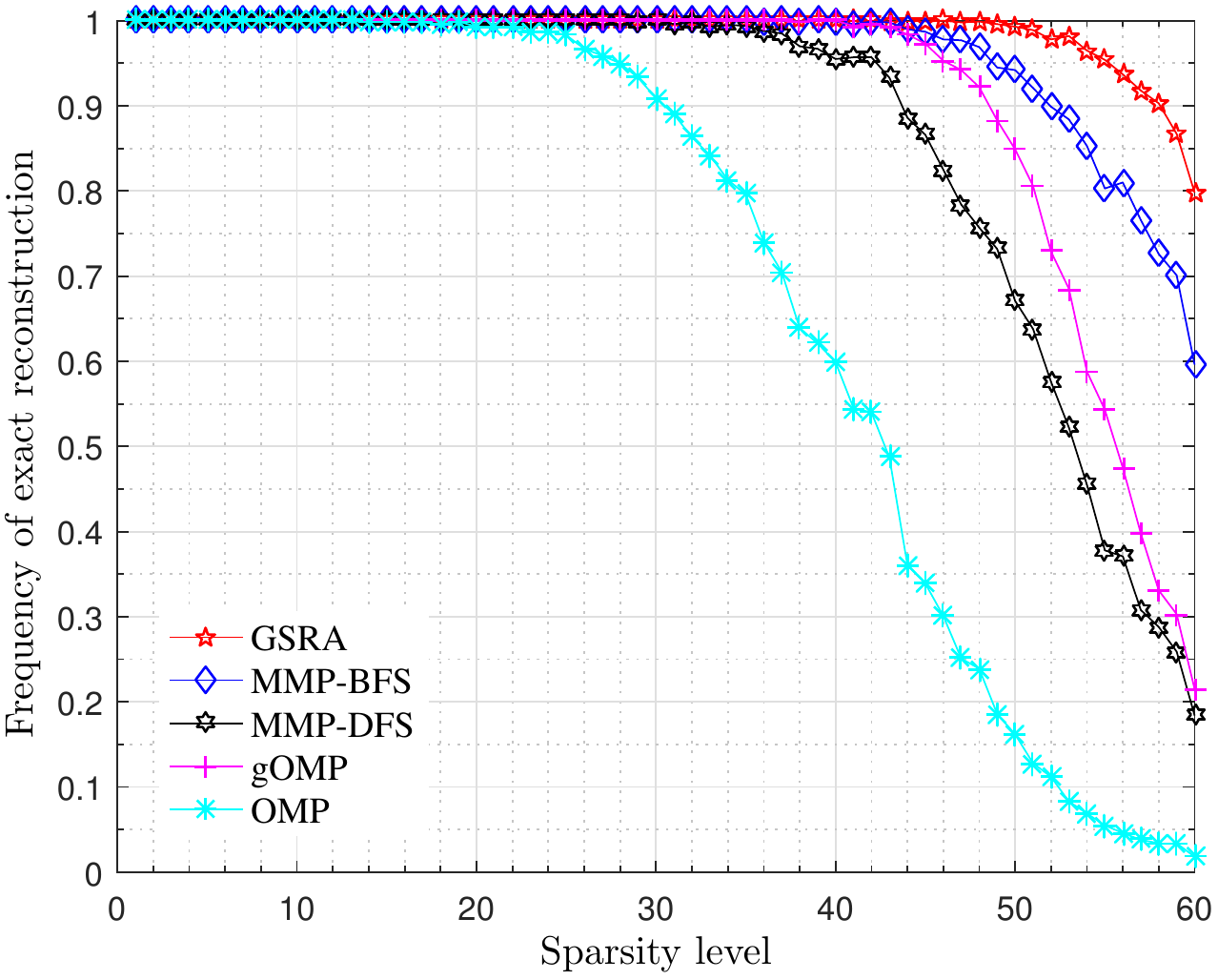}
a)
\end{minipage}
\begin{minipage}{\figwidth}
\centering
\includegraphics[width=\textwidth]{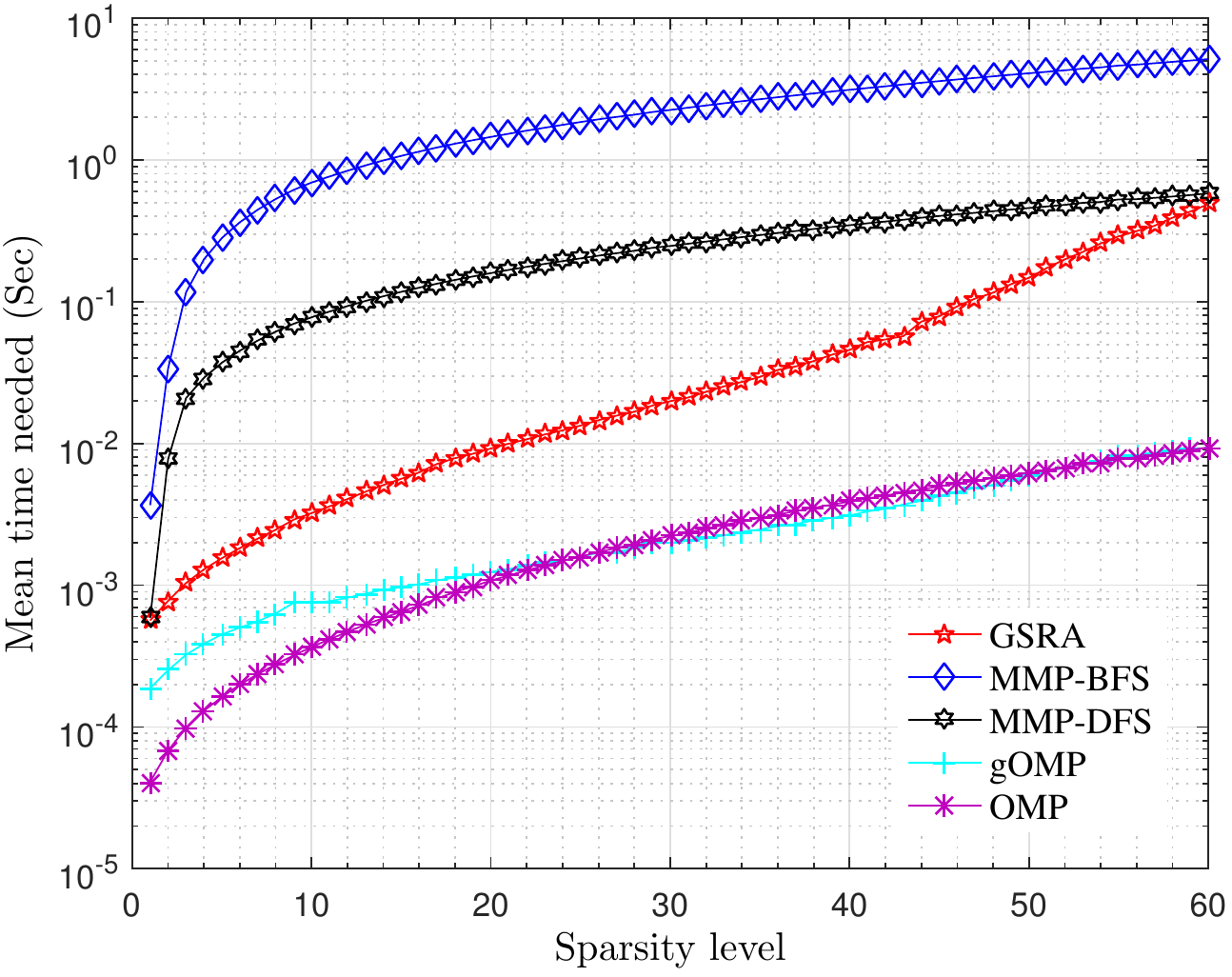}
b)
\end{minipage}
\caption{Performance comparison of reconstructing typical sparse signals in terms of two metrics:
a) frequency of exact reconstruction; b) mean time needed.}
\label{fig:Comparison}
\end{figure}

Figure~\ref{fig:Comparison} depicts the results of the above experiments. Figure~\ref{fig:Comparison}a) demonstrates that
GSRA can achieve an exact reconstruction with a greater frequency than OMP, gOMP, MMP-DFS and MMP-BFS for Gaussian signal of any sparsity level.
Concretely, OMP, gOMP, MMP-DFS and MMP-BFS can accurately reconstruct a signal of various sparsity levels, such as 21, 40, 35, and 43. In contrast, GSRA can still do this when the level reaches 50. Figure~\ref{fig:Comparison}b) shows that the time needed increases with an increase in the sparsity level, and the mean time needed by GSRA is shorter than that of MMP-DFS and MMP-BFS, but longer than that of OMP and gOMP.

\subsection{Reconstruction of 0--1 Sparse Signals}

In this experiment, we choose a $K$-subset of $\{1, 2, \cdots, n\}$, $K=1, 2, \cdots, 50$. For each value of $K$, we generate a signal $\vct{x}$ of sparsity level $K$, where the nonzero coefficients are set to one and those in the remaining indices to zero. The other configurations are the same as in the above sub-section.

\begin{figure}[!htb]
\centering
\begin{minipage}{\figwidth}
\centering
\includegraphics[width=\textwidth]{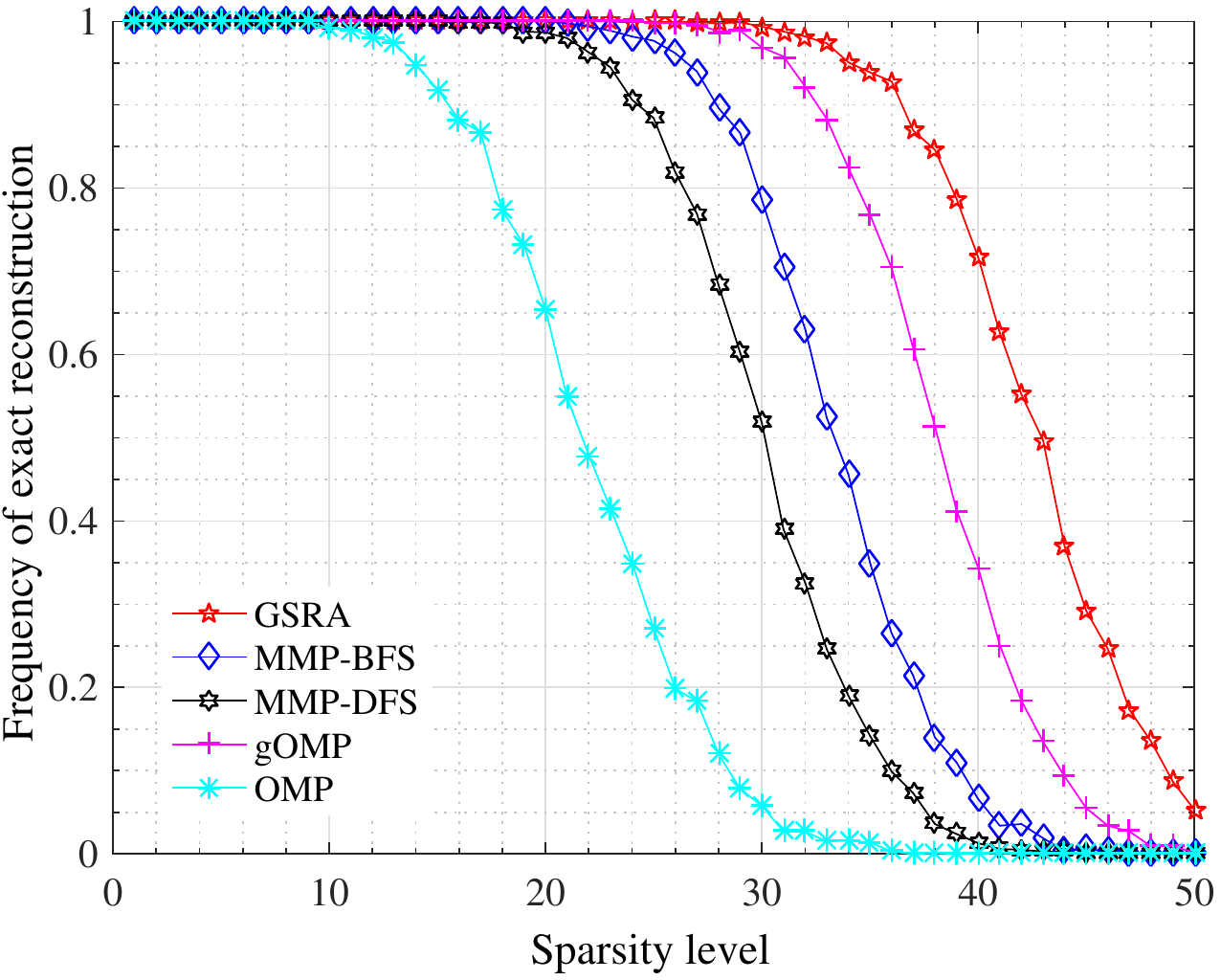}
a)
\centering
\includegraphics[width=\textwidth]{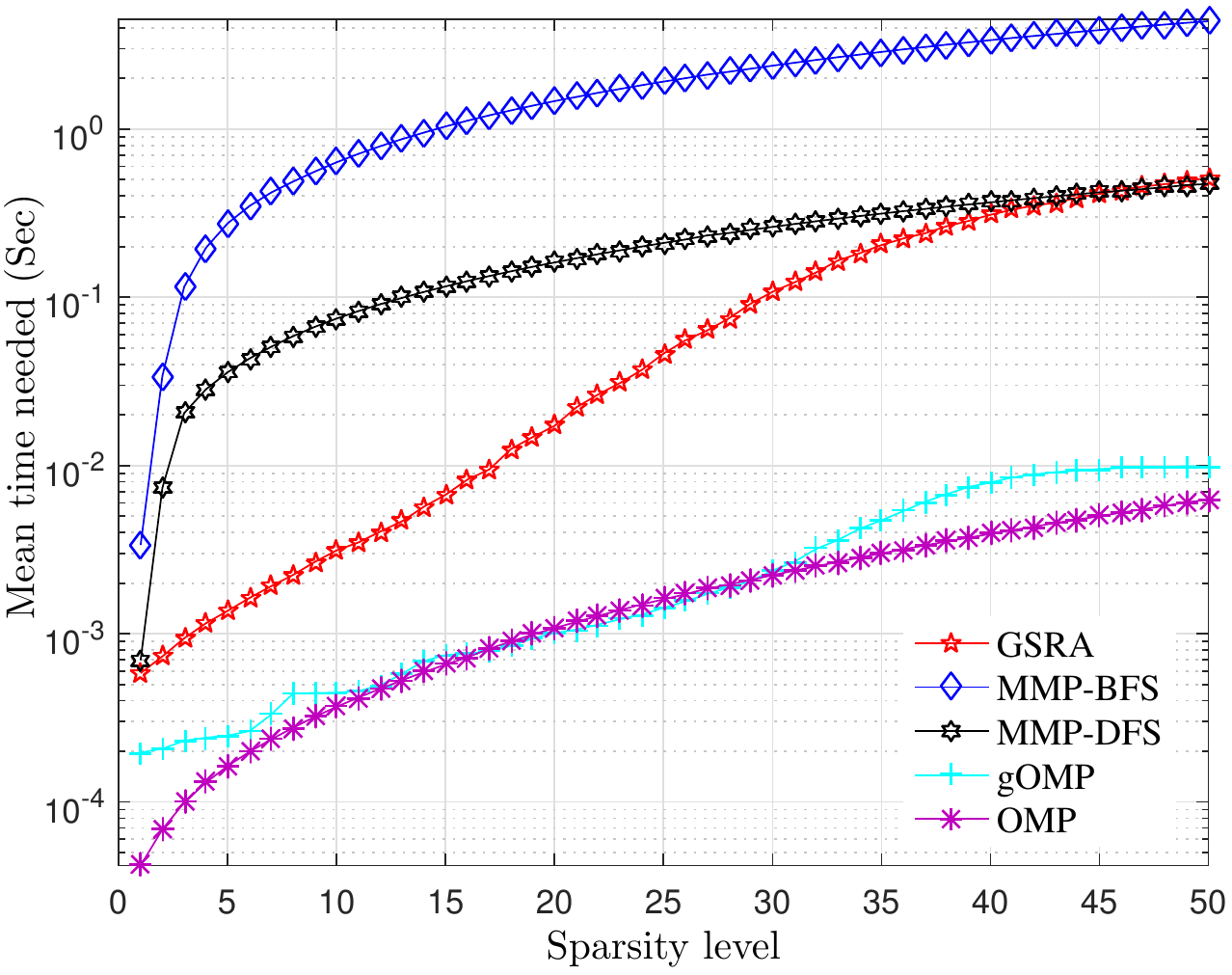}
b)
\end{minipage}
\begin{minipage}{\figwidth}

\end{minipage}
\caption{Performance comparison of reconstructing a 0--1 sparse signal in terms of two metrics:
a) frequency of exact reconstruction; b) mean time needed.}
\label{fig:Comp}
\end{figure}

Figure~\ref{fig:Comp}a) shows that the GSRA performs better than OMP, gOMP, MMP-DFS and MMP-BFS in terms of the frequency of exact reconstruction
for 0--1 sparse signals of any sparsity level. OMP, gOMP, MMP-DFS and MMP-BFS can accurately reconstruct signals of various sparsity levels: 10, 28, 20, and 23. Meanwhile, GSRA can obtain satisfying results when the sparsity level is increased to 30. As shown in Figure~\ref{fig:Comp}b), the comparison of the results
of GSRA with the other reference algorithms in terms of the mean time needed of GSRA are the same as that in the above subsection.

\subsection{Reconstruction of Synthetic Signals}

To further verify the performance of GSRA in comparison with the four reference algorithms, some experiments were done for the recovery
of an additive synthetic signal
\begin{equation}
\vct{y}=\Fee\vct{x}+\vct{w}.
\end{equation}
The synthesis degree is measured by the Signal to Measurement-Noise Ratio
\begin{equation}
\mathit{SMNR}\buildrel \Delta \over=10\log_{10} \frac{E \pnorm{2}{\vct{x}}^2}{E\pnorm{2}{\vct{w}}^2} = 10\log_{10} \frac {K\cdot \sigma_s^2}{m \cdot\sigma_l^2},
\label{eq:5}
\end{equation}
where $E(\bullet)$ denotes the mathematical expectation operator, $\sigma_s^2$ and $\sigma_l^2$ denote the power of each element of the signal and noise vector, respectively.
To measure the recovery degree more accurately, the Signal-to-Reconstruction-Error Ratio
\begin{equation*}
\mathit{SRER}\buildrel \Delta \over=10\log_{10} \frac{E \pnorm{2}{\vct{x}}^2 }{E\pnorm{2}{\vct{x}-\widehat{\vct{x}}}^2 }
\label{eq:4}
\end{equation*}
is adopted here instead.

The concrete steps of the experiment are described as follows.
\begin{enumerate}
\item Set the sparsity level $K=20$ and fix $n=500$. Then, adjust the value of the sampling rate $\varphi=m/n$ so that the number of measurements $m$ is an integer.

\item Generate elements of the $\Fee$ independently from a source following $\mathcal{N}(0, 1)$ and normalize each column norm to unity.

\item For the noisy region of the signal $\vct{x}$, the additive noise $\vct{w}$ is set as a Gaussian random vector whose elements are independently chosen
with distribution $\mathcal{N}(0, \sigma_w^2)$.

\item Select a measurement vector with $\mathit{SMNR}=20$.

\item Apply the reconstruction methods in Algorithm~\ref{algo:mp}.

\item Repeat steps 3)--5) $T=10$ times.

\item Repeat steps 2)--6) $Q = 100$ times.

\item Calculate the average value of the $\mathit{SRER}$ scores for the $T\cdot Q$ signals.
\end{enumerate}

As for GSRA, the other parameters are same as in Sec.~\ref{ssec:n01} except the search depth. With an increase of depth, the performance of GSRA is improved.
Typically, the search depth is set to be 16. The above eight experimental steps were executed with the sampling rate $\varphi$ ranging from 0.15 to 0.20,
and the obtained results are listed in Table~\ref{tab:mutual}, which demonstrates that GSRA performs better than any other reference scheme in each case.

\tabcolsep=3pt
\begin{table}[ht!]
\caption{Performance comparison of the recovery algorithms in terms of $\mathit{SRER}$ under some sampling rates.}
\begin{tabular}{*{6}{|c}|}
\hline
\backslashbox{ $\varphi$ }{Name} & OMP & gOMP & MMP-DFS & MMP-BFS & GSRA \\\hline
 0.15 & 8.11  & 7.75  & 15.74 & 15.38 & 16.05 \\  \hline
 0.16 & 9.57  & 9.08  & 18.09 & 17.63 & 18.30 \\  \hline
 0.17 & 12.24 & 11.37 & 18.47 & 17.51 & 19.07 \\  \hline
 0.18 & 14.71 & 13.48 & 19.67 & 18.51 & 19.20 \\  \hline
 0.19 & 17.49 & 15.08 & 20.28 & 19.43 & 20.39 \\  \hline
 0.20 & 19.31 & 15.90 & 20.89 & 20.04 & 20.98 \\ \hline
\end{tabular}
\label{tab:mutual}
\end{table}

\section{CONCLUSION}

To improve the accuracy of sparse signal recovery, this paper proposed an iterative greedy reconstruction algorithm by examining multiple promising candidates with the help of greedy search. The key feature of the algorithm is to use a hope-tree and the decreasing subspace pursuit method to obtain the final support set. Detailed experimental results demonstrated that the proposed algorithm performs well for Gaussian signals, 0--1 sparse signals, and synthetic signals. Research on making the depth of the GSRA adaptive deserves further exploration.

\bibliographystyle{IEEEtran}
\bibliography{secu_abrv,SPL_CS}

\end{document}